# Analysis of controlled Rabi flopping in a double rephasing photon echo scheme for quantum memories

**Rahmatullah[1,2] and B. S. Ham[1][†]**

[1]Center for Photon Information Processing, and School of Electrical Engineering and Computer Science, Gwangju Institute of Science and Technology, Gwangju, Republic of Korea
[2]Department of Physics, COMSATS Institute of Information Technology, Islamabad, Pakistan

[†]**E-mail:** bham@gist.ac.kr

**Abstract**
A double rephasing photon echo is analyzed for inversion-free photon echo-based quantum memories using controlled Rabi flopping, where the Rabi flopping is used for phase control of atom coherence. Unlike the rephasing-caused $\pi-$phase shift in a single rephasing scheme, the control Rabi flopping between the excited state and an auxiliary third state induces coherence inversion. Thus, the absorptive photon echo in a double rephasing scheme can be manipulated to be emissive via the controlled Rabi flopping. Here we present a quantum coherence control of atom phases in a double rephasing photon echo scheme for emissive photon echoes for quantum memory applications.

## 1. Introduction

Modified photon echoes have been intensively studied for quantum memory applications over the last decade since the first protocol of controlled reversible inhomogeneous broadening, where photon echo can be achieved by a counter-propagating control pulse set in a three-level Doppler [1] and non-Doppler [2] medium. Due to the inherent population inversion in photon echoes [3], resulting in quantum noises and violation of no cloning theorem [4], conventional photon echo itself cannot be directly applied to quantum memories. Compared with single atom-based quantum memory protocols, e.g., utilizing nuclear spins recently demonstrated in Si-based semiconductors [5], the photon echoes in rare-earth doped solids have benefits of multimode, ultrafast, and ultrahigh absorption [6]. To overcome the inherent population inversion in photon echoes, atomic frequency comb (AFC) echoes [7,8], gradient echoes [9,10], and controlled double rephasing (CDR) echoes [11,12] are presented for quantum memory applications. Because ultralong quantum memory is an essential component for long-distance quantum communications using quantum repeaters [13,14], a storage time extension has also been a critical issue [15-18]. As experimentally demonstrated by using dynamic decoupling (DD) [14] and optical locking via controlled coherence conversion (CCC) [19], the optical storage time can be extended up to spin population decay time.

The CCC theory has been proposed to convert an absorptive echo into an emissive one in a double rephasing (DR) photon echo scheme [11]. The DR photon echo scheme inherently gives the benefit of no population inversion. Because rephasing itself in photon echoes results in a time reversal process owing to a $\pi$ phase shift in collective atom coherence, the DR echo is obviously absorptive like the data pulse due to the $2\pi$ phase shift (no change) in coherence. Although silent echoes in doubly rephased echo schemes have been successfully demonstrated [20-22], the final echo extraction out of the medium is strongly prohibited due to its absorptive coherence [11,12]. Moreover, there is no way to solve this absorptive echo problem in a two-level system, at least not yet. However, DR photon echoes have been observed recently, which is seemingly violating the CDR echo theory. In the photon echo experiments, however, echoes are always observed regardless of the pulse area due to the imperfect rephasing-caused coherence leakage when commercial Gaussian lights are used. Recently calculated Gaussian pulse-caused echo efficiency is as high as 26% in the double rephasing scheme [23]. The CCC in CDR echoes has already been discussed in a single rephasing photon echo scheme theoretically [24] as well as experimentally [25].

Here in the present paper, we analytically investigate the collective atom phase control in a DR scheme, and confirm the CDR echo theory with a proof of coherence inversion. Compared with full numerical analysis in previous discussions [11,12,16,17,24], we present full analytic solutions in this article.

## 2. Theory

Figure 1 is a schematic diagram of the present CDR echoes, where the control pulse set of $C_1$ and $C_2$ is for the atom phase control in the DR scheme. The data (D), first rephasing ($R_1$) and second rephasing ($R_2$) pulses satisfy a double rephasing photon echo scheme, where they are resonant between states $|1\rangle$ and $|2\rangle$ as shown in Fig. 1(a). The pulse sequence of CDR is shown in Fig. 1(b), where the control pulse set $C_1$ and $C_2$ is resonant between states $|2\rangle$ and $|3\rangle$. The time delay $\tau$ between $C_1$ and $C_2$ is used for storage time extension, which is limited by the spin dephasing [25,26]. The spin dephasing can be minimized with the zero first-order Zeeman method [27]. In an optical locking scheme applied to three-pulse photon echoes [17], the storage time extends up to spin population decay time [19]. To satisfy general conditions of CDR, each pulse area of $R_1$, $R_2$, $C_1$, and $C_2$ is set to be $\pi$. The pulse area of D is set to small at $0.1\pi$. The pulse area is defined by $\varphi_i = \int \Omega_i dt$, and $\Omega_i (i = D, R_1, R_2, C_1$ and $C_2)$ is the Rabi frequency of the pulse.

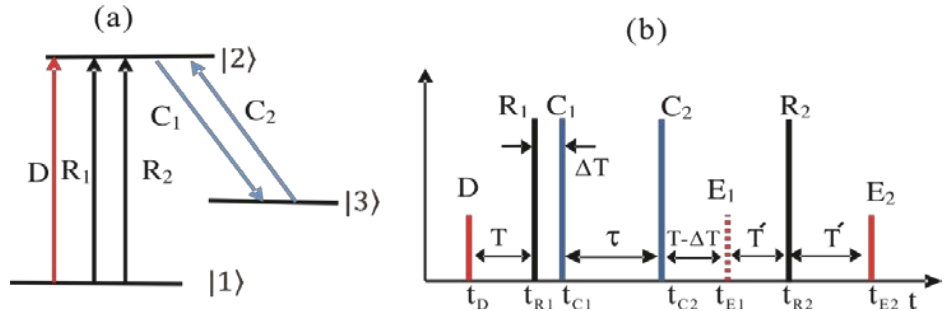

**Figure 1.** (a) Schematic of controlled double rephasing echoes. (b) Pulse sequence for (a). $t_j$ is the the arrival time of pulse j.

The CCC in CDR must be distinguished from resonant Raman or electromagnetically induced transparency (EIT) based on two-photon resonance without shelving on the excited state. For photon echo-based quantum memories, the signal (data) pulse information (phase and amplitude) must be fully transferred into a matter (optical coherence) state via a complete absorption process in an optically dense, inhomogeneously broadened two-level medium [16]. Unlike other coherence optics in a three-level system mentioned above, the inhomogeneity of the ensemble is the fundamental requirement for the coherence evolutions in photon echoes. One unique property of the CCC is the double coherence swapping between the optical and spin states via the control pulse set of $C_1$ and $C_2$. Unlike EIT, The $R_1$ and $C_1$ must be avoided from the two-photon Raman coherence, where the delay $\Delta T$ between $R_1$ and $C_1$ must be longer than the inverse of inhomogeneous width. Usually this requirement is easily satisfied even for the consecutive $\pi$-optical pulse sequence in most rare-earth doped solids [25].

The purpose of $C_1$ is simply to hold both optical phase decay and optical phase evolutions via complete population transfer from the excited state $|2\rangle$ to the auxiliary spin state $|3\rangle$, resulting in optical coherence $\rho_{12}=0$ [11]. For this the state $|3\rangle$ must be vacant initially. When the second control pulse $C_2$ is turned on, the system population is completely recovered to the initial one reached by $R_1$. However, the system coherence is not invariant, resulting in absorptive photon echo $E_2$ [12,24].

The interaction picture Hamiltonian in the atom-field interactions under rotating-wave approximation of the proposed system in Fig. 1(a) is written as:

$$H = -\hbar/2 \begin{bmatrix} 0 & \Omega_j & 0 \\ \Omega_j & 0 & \Omega_k \\ 0 & \Omega_k & 0 \end{bmatrix}, \tag{1}$$

where, $\Omega_j$ ($j$= D, $R_1$ or $R_2$) is the Rabi frequency of D, $R_1$ and $R_2$, or $\Omega_k$ ($k$= $C_1$ or $C_2$) is the Rabi frequency of $C_1$ or $C_2$. We calculate the rate equations for the density matrix elements using Von Neumann equation [28]:

$$\dot{\rho} = -\frac{i}{\hbar}[H,\rho] - \frac{1}{2}\{\Gamma,\rho\}. \tag{2}$$

The corresponding rate equations are

$$\dot{\rho}_{11} = -i\frac{\Omega_j}{2}(\rho_{12} - \rho_{21}), \tag{3a}$$

$$\dot{\rho}_{22} = -i\frac{\Omega_j}{2}(\rho_{21} - \rho_{12}) - i\frac{\Omega_k}{2}(\rho_{23} - \rho_{32}), \tag{3b}$$

$$\dot{\rho}_{33} = -i\frac{\Omega_k}{2}(\rho_{32} - \rho_{23}), \tag{3c}$$

$$\dot{\rho}_{12} = -i\frac{\Omega_j}{2}(\rho_{11} - \rho_{22}) - i\frac{\Omega_k}{2}\rho_{13}, \tag{3d}$$

$$\dot{\rho}_{12} = -i\frac{\Omega_k}{2}\rho_{12} + i\frac{\Omega_j}{2}\rho_{23}, \tag{3e}$$

$$\dot{\rho}_{23} = -i\frac{\Omega_k}{2}(\rho_{22} - \rho_{33}) + i\frac{\Omega_j}{2}\rho_{13}, \tag{3f}$$

where, all decay rates are set to be zero for simplicity. Thus the the overall dephasing is only due to the atom detuning in the inhomogeneous broadening. We now consider the CDR echo scheme for the following discussion. For this we start with a general DR scheme to investigate the absorptive coherence on the final echo E2 without C1 and C2 pulses in Fig. 1.

### 3. Discussion
### 3.1. DR photon echoes
In this sub-section, we study conventional two-pulse photon echoes in a DR scheme without $C_1$ and $C_2$ in Fig. 1. We derive time-dependent density matrix equations for the expressions of ensemble coherence between the ground and excited states, and population in each bare state.

### 3.1.1. D-pulse
We first derive the expressions of coherence and population excited by the D-pulse. The equations of motion for D-pulse by setting $\Omega_j = \Omega_D$ and $\Omega_k = 0$ in equation (3) are as follows:

$$\dot{\rho}_{11} = -i\frac{\Omega_D}{2}(\rho_{12} - \rho_{21}), \tag{4a}$$

$$\dot{\rho}_{22} = -i\frac{\Omega_D}{2}(\rho_{21} - \rho_{12}), \tag{4b}$$

$$\dot{\rho}_{12} = -i\frac{\Omega_D}{2}(\rho_{11} - \rho_{22}), \tag{4c}$$

$$\dot{\rho}_{21} = -i\frac{\Omega_D}{2}(\rho_{22} - \rho_{11}). \tag{4d}$$

Initially all atoms are in the ground state $|1\rangle$: $\rho_{11}(0) = 1$; $\rho_{22}(0) = \rho_{12}(0) = \rho_{21}(0) = 0$. The Laplace transform of equation (4) with $\rho_{11} + \rho_{22} = 1$ yields:

$$\mathcal{L}[\rho_{11}] = \frac{2s^2 + \Omega_D^2}{2s(s^2 + \Omega_D^2)}, \tag{5a}$$

$$\mathcal{L}[\rho_{12}] = \frac{-i\Omega_D}{(s^2 + \Omega_D^2)}, \tag{5b}$$

$$\mathcal{L}[\rho_{21}] = \frac{i\Omega_D}{(s^2 + \Omega_D^2)}. \tag{5c}$$

The final equations for population and coherence are obtained by taking the inverse Laplace transform of equation (5):

$$\rho_{11} = \cos^2\left(\frac{\varphi_D}{2}\right), \tag{6a}$$

$$\rho_{22} = \sin^2\left(\frac{\varphi_D}{2}\right), \tag{6b}$$

$$\rho_{12} = -\frac{i}{2}\sin(\varphi_D), \tag{6c}$$

where $\varphi_D$ is the area of the D-pulse. The D-pulse obeys the area theorem which has a direct relation with coherence [29]:

$$\frac{\partial \varphi_D}{\partial z} = -\frac{\alpha}{2}\sin(\varphi_D), \tag{7}$$

where $\alpha$ is the absorption coefficient. For the D-pulse with a very small pulse area, $\sin(\varphi_D) \approx 1$, $\varphi_D = (\varphi_D)_0 e^{-\alpha z/2}$ represents the Beer's law. The information of D-pulse is now transferred into the atom coherence. For a weak D-pulse, $\varphi_D \ll 1$, the atomic population still remains in the ground state $|1\rangle$: $\rho_{11} \approx 1$; $\rho_{22} \approx 0$. In our analysis, the D-pulse area is set to be $0.1\pi$.

### 3.1.2. $R_1$-pulse

As soon as the atoms are excited by D, they immediately start to evolve with their own detuning-dependent phase velocity until the rephasing pulse ($R_1$-pulse) comes. We use equation (6) as initial conditions for $\Omega_j = \Omega_{R_1}$ and $\Omega_k = 0$ to solve equation (3). The solution of the rate equations for $R_1$-pulse is as followings:

$$\rho_{11} = \cos^2\left(\frac{\varphi_D + \varphi_{R_1}}{2}\right), \tag{8a}$$

$$\rho_{22} = \sin^2\left(\frac{\varphi_D + \varphi_{R_1}}{2}\right), \tag{8b}$$

$$\rho_{12} = -\frac{i}{2}\sin(\varphi_D + \varphi_{R_1}). \tag{8c}$$

Equation (8c) indicates that the rephasing $\pi$ pulse $R_1$ results in a $\pi$ phase shift in the coherence $\rho_{12}$ initiated by the D-pulse in equation (6c), as shown in Fig. 2(a): $[\rho_{12}] \xrightarrow{R1} [\rho_{12}]^*$ (see also AppendixA for $\pi/2$ pulse area of D). All details of detuning-dependent atom phase evolutions and rephasing effects are numerically shown in Fig. 4 of ref. [24], where the real part of $\rho_{12}$ is exactly symmetric, cancelling coherence each other. The $\pi$ rephasing pulse swaps the population between ground and excited states as shown in Fig. 2(b), resulting in spontaneous and/or stimulated emission noises. To overcome the population inversion, a controlled double rephasing concept has been developed in the name of CDR echoes [11,12]. For DR echoes, the second $\pi$ optical pulse $R_2$ is added to swap the populations again, where the second echo $E_2$ is free from quantum noises. To fully restore the D-pulse transferred coherence, the first echo $E_1$ must be erased (or silenced), where this erasing process does not affect the individual coherence evolutions [20-22].

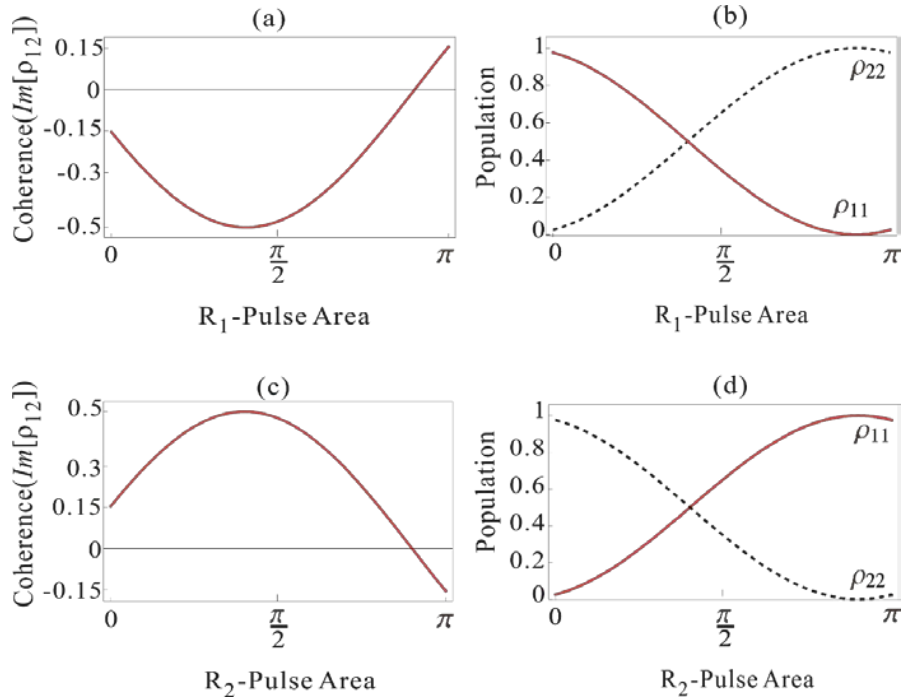

**Figure 2.** (a) Plot of Im[$\rho_{12}$] (equation 8c) versus $R_1$-pulse area $\varphi_{R_1}$ with area of D-pulse $\varphi_D = 0.1\pi$. (b) Corresponding population evolution (red) $\rho_{11}$ (equation 8a) and (dotted) $\rho_{22}$ (equation 8b). (c) Plot of Im[$\rho_{12}$] (equation 9c) versus $R_2$-pulse area $\varphi_{R_2}$ with area of D-pulse $\varphi_D = 0.1\pi$ and $R_1$ is $\varphi_{R_1} = \pi$. (d) Corresponding population evolution (red) $\rho_{11}$ (equation 9a) and dotted $\rho_{22}$ (equation 9b).

We derive the coherence and population rate equations for $R_2$-pulse by replacing $\Omega_j$ with $\Omega_{R_2}$ and setting $\Omega_k = 0$ in equation (3). We use equation (8) as initial conditions and calculate the expressions for coherence and populations as follows:

$$\rho_{11} = \cos^2\left(\frac{\varphi_D + \varphi_{R_1} + \varphi_{R_2}}{2}\right), \tag{9a}$$

$$\rho_{22} = \sin^2\left(\frac{\varphi_D + \varphi_{R_1} + \varphi_{R_2}}{2}\right), \tag{9b}$$

$$\rho_{12} = -\frac{i}{2}\sin(\varphi_D + \varphi_{R_1} + \varphi_{R_2}). \tag{9c}$$

In Figs. 2(c) and (d), the $R_2$ pulse area-dependent coherence and population are shown for $\varphi_D = 0.1\pi$ and $\varphi_{R_1} = \pi$. As shown in Fig. 2(c), the $\pi$–$R_2$ pulse inverts the coherence as $\pi$–$R_1$ pulse does. Here, the negative sign in the coherence $\rho_{12}$ shows absorption. Thus the second echo by $R_2$ is absorptive like the data pulse D [11,12]. This means that the generated echo $E_2$ in the DR scheme cannot be radiated out of the medium, as D is fully absorbed into the medium. By the way, the observations of $E_2$ in refs. [20-22] have been understood as imperfect rephasing-caused coherence leakage due to Gaussian distributed light pulses [23]. Our aim is here, to get the inversion-free emissive echo. To convert the absorptive echo $E_2$ in Fig. 2 into an emissive one, the CDR echo scheme is applied. In the following section, we describe the role of $C_1$ and $C_2$ for CCC in details.

### 3.2 CDR echoes

In this sub-section, we discuss the CDR echo of Fig. 1 by inserting the control pulse set of $C_1$ and $C_2$ in the DR scheme. The control pulse set can follow either $R_1$ as shown in Fig. 1(b) or $R_2$ as discussed in refs. [11,12]. In both cases, $C_1$ must be activated before the echo timing [24].

### 3.2.1. $C_1$-pulse

The function of $C_1$-pulse with a $\pi$ pulse area is to temporally hold optical coherence decay as well as optical phase evolution via transferring population in the excited state $|2\rangle$ to an auxiliary ground (spin) state $|3\rangle$. In general, spin phase decay time is much longer than the optical counterpart in rare-earth doped crystals. Thus, C1 plays a role of storage time extension [2,25,26]. The coherence and population changes by $C_1$ can be obtained by using equation (8) as initial conditions. The solutions of density matrix equation (3) for $C_1$ are obtained as:

$$\rho_{11} = \cos^2\left(\frac{\varphi_D + \varphi_{R_1}}{2}\right), \tag{10a}$$

$$\rho_{22} = \cos^2\left(\frac{\varphi_{C_1}}{2}\right)\sin^2\left(\frac{\varphi_D + \varphi_{R_1}}{2}\right), \tag{10b}$$

$$\rho_{33} = \sin^2\left(\frac{\varphi_{C_1}}{2}\right)\sin^2\left(\frac{\varphi_D + \varphi_{R_1}}{2}\right), \tag{10c}$$

$$\rho_{12} = -\frac{i}{2}\cos\left(\frac{\varphi_{C_1}}{2}\right)\sin(\varphi_D + \varphi_{R_1}), \tag{10d}$$

$$\rho_{13} = -\frac{1}{2}\sin\left(\frac{\varphi_{C_1}}{2}\right)\sin(\varphi_D + \varphi_{R_1}), \tag{10e}$$

$$\rho_{23} = -\frac{i}{2}\sin(\varphi_{C_1})\sin^2\left(\frac{\varphi_D + \varphi_{R_1}}{2}\right). \tag{10f}$$

The optical coherence $\rho_{12}$ in equation (10d) by $C_1$-pulse is equal to $\cos(\varphi_{C_1}/2)$ times the coherence generated by $R_1$-pulse in equation (8c), where the $R_1$-resulted coherence is 0.15 for the $0.1\pi$ of D-pulse and $\pi$ of $R_1$-pulse (see Fig. 3). So equation (10d) becomes $\rho_{12} = 0.15i\cos(\varphi_{C_1}/2)$ (see also Fig. 2(a)). Similarly the spin coherence in equatioin (10-e) is $\rho_{13} = 0.15\sin(\varphi_{C_1}/2)$. In the absence of the $C_1$-pulse, i.e., $\varphi_{C_1} = 0$, $\rho_{12} = 0.15i$ and $\rho_{13} = 0$. In the presence of the $\pi$ $C_1$-pulse, the optical and spin coherence becomes $\rho_{12} = 0.15i\cos(\pi/2) = 0$ and $\rho_{13} = 0.15\sin(\pi/2) = 0.15ie^{-i\pi/2}$, respectively. Thus, the $\pi$–$C_1$ pulse adds a $\pi/2$ phase shift to the transferred coherence $\rho_{13}$ [28]. This is a well-known property in a resonant two filed interactions in a three-level atomic system, where there is a π/2 phase shift between $Im\rho_{12}$ and $Re\rho_{13}$. In conclusion, $C_1$ pulse locks both optical phase decay and coherence evolutions, while transfers $\rho_{12}$ into $\rho_{13}$ with a π/2 phase shift via complete population transfer. In other words $Im\rho_{12}$ becomes $Re\rho_{13}$ as shown in Fig. 3(a). Here $Im\rho_{13}$ is zero as $Re\rho_{12}$ is zero in equation (6).

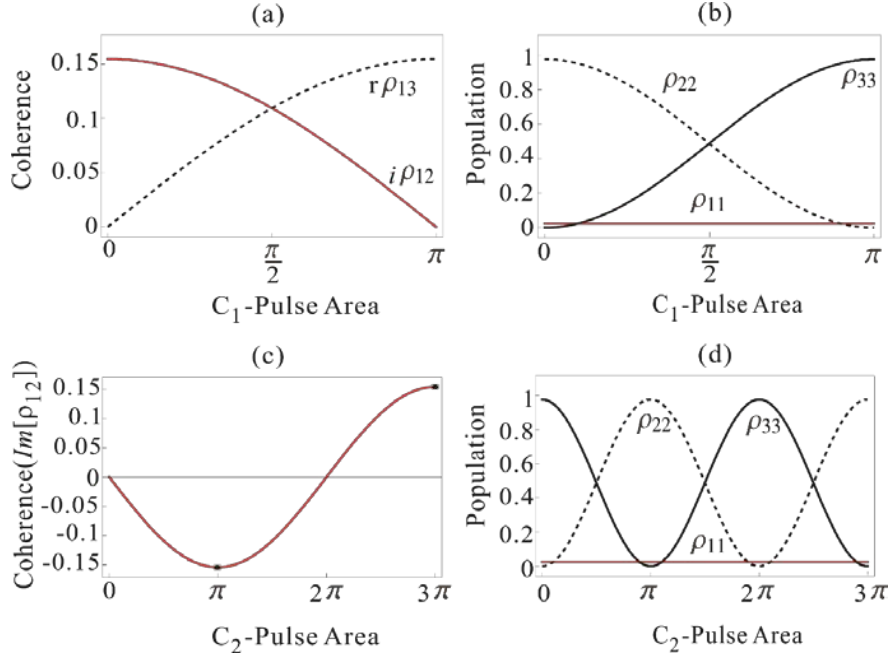

**Figure 3.** (a) Plot of $Im[\rho_{12}]$ and $Re[\rho_{13}]$ (equation 10d and 10e) versus $C_1$-pulse area $\varphi_{C_1}$ with area of D-pulse $\varphi_D = 0.1\pi$ and $R_1$ is $\varphi_{R_1} = \pi$. (b) Corresponding population evolution (red) $\rho_{11}$ (equation 10a), (dotted) $\rho_{22}$ (equation 10b) and (black) $\rho_{33}$ (equation 10c). (c) Plot of $Im[\rho_{12}]$ (equation 11d) versus $C_2$-pulse area $\varphi_{C_2}$ with area of the other pulses are $\varphi_D = 0.1$, $\varphi_{R_1} = \pi$ and $\varphi_{C_1} = \pi$. (d) Corresponding population evolution (red) $\rho_{11}$ (equation 11a), (dotted) $\rho_{22}$ (equation 11b) and (black) $\rho_{33}$ (equation 11c).

### 3.2.2. $C_2$-pulse
The function of $C_2$-pulse is to restore the transferred coherence by C1. Using equation (10) as initial conditions, and by setting $\Omega_k = \Omega_{C_2}$ and $\Omega_j = 0$ in equation (3), the system coherence and population expressions by $C_2$-pulse are obtained as follows:

$$\rho_{11} = \cos^2\left(\frac{\varphi_D + \varphi_{R_1}}{2}\right), \tag{11a}$$

$$\rho_{22} = \cos^2\left(\frac{\varphi_{C_1} + \varphi_{C_2}}{2}\right)\sin^2\left(\frac{\varphi_D + \varphi_{R_1}}{2}\right), \tag{11b}$$

$$\rho_{33} = \sin^2\left(\frac{\varphi_{C_1} + \varphi_{C_2}}{2}\right)\sin^2\left(\frac{\varphi_D + \varphi_{R_1}}{2}\right), \tag{11c}$$

$$\rho_{12} = -\frac{i}{2}\cos\left(\frac{\varphi_{C_1} + \varphi_{C_2}}{2}\right)\sin(\varphi_D + \varphi_{R_1}), \tag{11d}$$

$$\rho_{13} = -\frac{1}{2}\sin\left(\frac{\varphi_{C_1} + \varphi_{C_2}}{2}\right)\sin(\varphi_D + \varphi_{R_1}), \tag{11e}$$

$$\rho_{23} = -\frac{i}{2}\sin(\varphi_{C_1} + \varphi_{C_2})\sin^2\left(\frac{\varphi_D + \varphi_{R_1}}{2}\right). \tag{11f}$$

The coherence in equation (11d) is equal to $\cos\big((\varphi_{C_1} + \varphi_{C_2})/2\big)$ multiply by the coherence excited by $R_1$-pulse in equation (8c). The $\pi$-$\pi$ pulse sequence of $C_1$ and $C_2$, therefore, induces a coherence inversion via the round trip of population transfer between the excited and auxiliary states: $\cos((\pi + \pi)/2) = -1$ (see Fig. 3(c) [11,12,24]: $\rho_{12} \xrightarrow{C_1 \& C_2} -\rho_{12}$. This coherence inversion mechanism is completeley different from the rephasing by R1 or R2 [12]. In order to resume the coherence initiated by the $R_1$-pulse, the sum pulse area of $C_1$ and $C_2$ must be equal to $4n\pi$ ($n = 1,2,3 ...$). In Figs. 3(c) and 3(d), we plot the coherence and population as a function of the $C_2$-pulse area for $\varphi_D = 0.1\pi$, $\varphi_{R_1} = \pi$ and $\varphi_{C_1} = \pi$. Figure 3(c) shows that the ensemble coherence excited by D and rephased by $R_1$ is recovered with $3\pi$ $C_2$-pulse. The $3\pi$ $C_2$, of course, returns the population from state $|3\rangle$ to the excited state $|2\rangle$ as shown in Fig. 3(d). Thus, the ($\pi$-$\pi$) $C_1$-$C_2$ pulse sequence in controlled AFC [26] induces an absorptive echo as in the DR scheme in Fig. 2(c) due to the $\pi$ phase shift by the control Rabi flopping. The experimental observation in ref. [26] is not an artifact but due to the coherence leakage by imperfect rephasing by commercial Gaussian distributed laser pulses, where its maximum echo efficiency is far less then unity [23]. In the CDR echo scheme, however $\pi$–$\pi$ pulse sequence of C1 and C2 is required to compensate the $\pi$ phase shift in the DR scheme.

### 3.2.3. $R_2$-pulse
For the CDR echo in Fig. 1, the final analytic solutions of density matrix equation (3) are obtained by using equation (11) as initial conditions:

$$\rho_{11} = \frac{1}{16}\bigg[\cos\left(\frac{\varphi_{C_1}+\varphi_{C_2}-\varphi_D-\varphi_{R_2}-\varphi_{R_1}}{2}\right) - \cos\left(\frac{\varphi_{C_1}+\varphi_{C_2}-\varphi_D+\varphi_{R_2}-\varphi_{R_1}}{2}\right) + 2\cos\left(\frac{\varphi_D-\varphi_{R_2}+\varphi_{R_1}}{2}\right) - \cos\left(\frac{\varphi_{C_1}+\varphi_{C_2}+\varphi_D-\varphi_{R_2}+\varphi_{R_1}}{2}\right) + \cos\left(\frac{\varphi_{C_1}+\varphi_{C_2}+\varphi_D+\varphi_{R_2}-\varphi_{R_1}}{2}\right) + 2\cos\left(\frac{\varphi_D+\varphi_{R_2}+\varphi_{R_1}}{2}\right)\bigg]^2, \tag{12a}$$

$$\rho_{22} = \frac{1}{16}\bigg[\sin\left(\frac{\varphi_{C_1}+\varphi_{C_2}-\varphi_D-\varphi_{R_2}-\varphi_{R_1}}{2}\right) + \sin\left(\frac{\varphi_{C_1}+\varphi_{C_2}-\varphi_D+\varphi_{R_2}-\varphi_{R_1}}{2}\right) + 2\sin\left(\frac{\varphi_D-\varphi_{R_2}+\varphi_{R_1}}{2}\right) - \sin\left(\frac{\varphi_{C_1}+\varphi_{C_2}+\varphi_D-\varphi_{R_2}+\varphi_{R_1}}{2}\right) - \sin\left(\frac{\varphi_{C_1}+\varphi_{C_2}+\varphi_D+\varphi_{R_2}+\varphi_{R_1}}{2}\right) - 2\sin\left(\frac{\varphi_D+\varphi_{R_2}+\varphi_{R_1}}{2}\right)\bigg]^2, \tag{12b}$$

$$\rho_{33} = \sin^2\left(\frac{\varphi_{C_1}+\varphi_{C_2}}{2}\right)\sin^2\left(\frac{\varphi_D+\varphi_{R_1}}{2}\right), \tag{12c}$$

$$\rho_{12} = -\frac{i}{16}\bigg[2\sin(\varphi_{R_2}) + 2\sin(\varphi_{R_2})\cos(\varphi_D + \varphi_{R_1})\left(3 + \cos\left(\frac{\varphi_{C_1}+\varphi_{C_2}}{2}\right)\right) + \sin(\varphi_{C_1} + \varphi_{C_2} - \varphi_{R_2}) - \sin(\varphi_{C_1} + \varphi_{C_2} + \varphi_{R_2}) + 8\cos(\varphi_{R_2})\sin(\varphi_D + \varphi_{R_1})\cos\left(\frac{\varphi_{C_1}+\varphi_{C_2}}{2}\right)\bigg]. \tag{12d}$$

In Fig. 4, we plot the evolutions of coherence and population as a function of $R_2$-pulse area for $\varphi_D = 0.1\pi$, $\varphi_{R_1} = \pi$, $\varphi_{C_1} = \pi$ and $\varphi_{C_2} = \pi$. As a result, both coherence and population excited by D are recovered with a $\pi$ pulse area of $R_2$, where spontaneous and stimulated emission-caused quantum noises are completely eliminated.

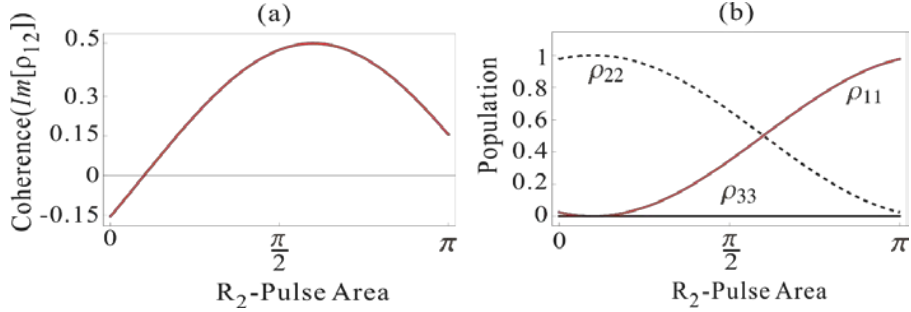

**Figure 4.** (a) Plot of Im[$\rho_{12}$] (equation 12d) versus $R_2$-pulse area $\varphi_{R_2}$. The area of other pulses are $\varphi_D = 0.1$, $\varphi_{R_1} = \pi$, $\varphi_{C_1} = \pi$ and $\varphi_{R_2} = \pi$. (b) Corresponding population evolution (red) $\rho_{11}$ (equation 12a), (dotted) $\rho_{22}$ (equation 12b) and (black) $\rho_{33}$ (equation 12c).

The present scheme can experimentally be realized in a rare-earth $Pr^{3+}$-doped $Y_2SiO_5$. In most rare-earth doped media, the ground state hyperfine splitting is a few tens of magahertz. Thus, tens of GHz of optical inhomogeneous broadening can be sliced into many spectral channels for multiple quantum memories, where practical parameter of optical Rabi frequency is ~MHz. For an extended storage time by $C_1$, Zeeman states may be used [29].

In an atomic ensemble such as Rb vapors, Zeeman splitting may also be used, where optical polarization control has been adapted to form a three-level system. However, such an atomic medium may not a good candidate for the photon echo-based quantum memory applications simply due to fast atomic diffusion. Moreover, providing a $\pi$ optical pulse in a few ns pulse duration within the optical phase decay time is very challenging with a commercial cw laser system.

## 4 Conclusion

In conclusion, we analytically presented the CDR echo protocol for spontaneous emission-free-photon echo-based quantum memory applications by combining double rephasing photon echoes with control Rabi flopping. For this, time-dependent density matrix equations were analytically solved for coherence/population evolutions to investigate the phase shift of a resonant atom. To overcome the absorptive echo problem in a bare double rephasing photon echo scheme, consecutive π−π control pulse sequence is inserted right after the first rephasing pulse. The control pulse-generated π phase shift was exactly compensated with another π phase shift resulted from the double rephasing scheme. As a result emissive photon echoes were obtained under no population inversion.

## Acknowledgment

This work was supported by the ICT R&D program of MSIP/IITP (1711028311: Reliable crypto-system standards and core technology development for secure quantum key distribution network.

## APPENDIX A: Coherence swapping using $\pi/2$ D-pulse

Here we analyze the coherence swapping by the π control pulse C1 for a $\pi/2$ D-pulse in Fig. 3. The π/2 D-pulse creates maximum coherence between $|1\rangle \leftrightarrow |2\rangle$ transition, where $Im[\rho_{12}] = -0.5$ (see equation (6c) for D= $\pi/2$). The negative sign represents absorption of the D-pulse. In Fig. 5, we show a phase shift of the D-pulse excited coherence according to each applied pulses. The first rephasing pulse $R_1$ switches the coherence from absorption to emission as shown in Fig. 5(a).

Identical $C_1$ and $C_2$ pulses with π pulse area each invert the sign of coherence obtained by $R_1$, resulting in an absorptive echo in Figs. 5(b) and (c). Finally, the second rephaisng pulse $R_2$ adds a π phase shift to make the emissive photon echo under no population inversion as shown in Fig. 5(d). Either control pulse set or double rephasing pulse set with a 2π pulse area each does not change the D-excited population distribution. Thus, the CDR echo is confirmed for a noise free quantum memory protocol.

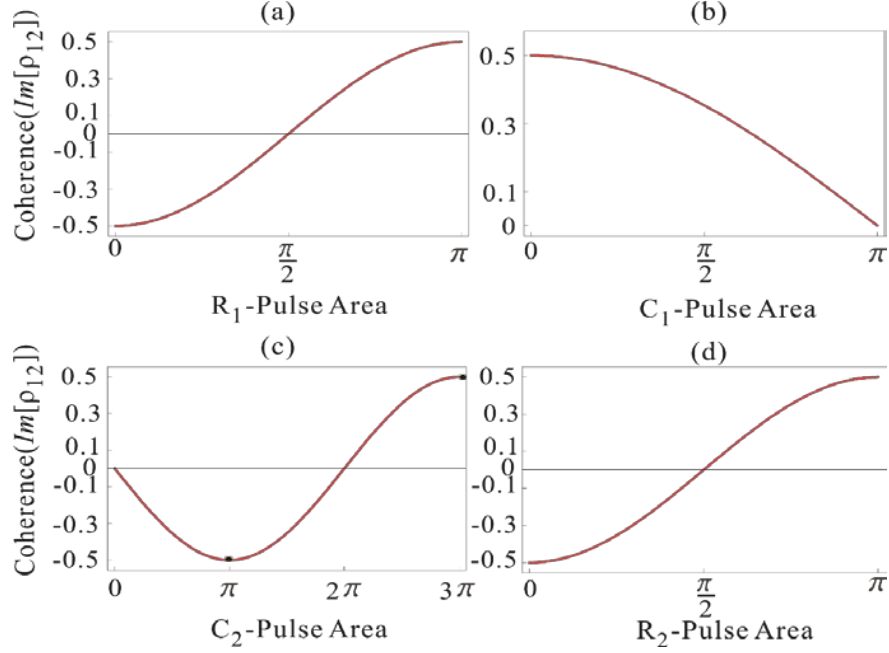

**Figure 5.** (a) Plot of $Im[\rho_{12}]$ versus (a) $R_1$-pulse area $\varphi_{R_1}$ (equation 8c), (b) $C_1$-pulse area $\varphi_{C_1}$ (equation 10d), (c) $C_2$-pulse area $\varphi_{C_2}$ (equation 11d) and $R_2$-pulse area $\varphi_{R_2}$ (equation 12d). The area of D-pulse is $\varphi_D = \pi/2$.